\documentclass[11pt,a4paper]{JHEP3}
\pdfoutput=1
\usepackage{epsfig}

\title{Berry's connection, K\"{a}hler geometry and the Nahm construction of monopoles}

\author{Kenny Wong\\
Department of Applied Mathematics and Theoretical Physics, \\
Centre for Mathematical Sciences, \\
University of Cambridge, \\
Cambridge, CB3 0WA, UK\\{\tt k.wong@damtp.cam.ac.uk}}

\abstract{We study supersymmetric deformations of $\mathcal N = 4$ quantum mechanics with a K\"{a}hler target space admitting a holomorphic isometry. We show that the twisted mass deformation generalises to a deformation constructed from matrix-valued functions of the moment map, which obey the Nahm equations. We also explain how $\mathcal N = 4$ supersymmetry implies that the Berry connection on the vacuum bundle for this theory satisfies the BPS monopole equations. In the case where the target space is a  Riemann sphere, our analysis reduces to the standard Nahm construction of monopoles. This generalises an earlier result by Sonner and Tong to the case of monopoles of magnetic charge greater than one.}

\begin{document}
\maketitle
\flushbottom
\newpage

\section*{Introduction}

\paragraph{}
A Berry phase \cite{berry} is a shift in the phase of a quantum-mechanical wavefunction due to adiabatic evolution of parameters in the Hamiltonian. Berry phases are encoded in a geometric object: a connection on a vector bundle over the parameter space. A well-known system that exhibits a Berry phase is the spin-half particle in an external magnetic field; here, the Berry connection is simply the Dirac monopole.

\paragraph{}
In \cite{sonner1}, Sonner and Tong found a similar setup in which the Berry connection that arises is a single \emph{'t Hooft-Polyakov} monopole \cite{thooft, polyakov}. The construction of Sonner and Tong uses $\mathcal N = 4$ supersymmetric quantum mechanics on a $\mathbb{CP}^1$ target space, where the background parameters being varied are twisted masses associated with an isometry of $\mathbb{CP}^1$.  Later, in \cite{sonner2}, Sonner and Tong showed that the fact that their Berry connection obeys the Bogomolny-Prasad-Sommerfield (BPS) monopole equations \cite{bogomolny, ps} is a consequence of supersymmetry, via an argument similar to Cecotti and Vafa's derivation of the tt* equations \cite{cecotti}, or the calculation of the effective action on the Coulomb branch for gauged linear models \cite{smilga, denef}.

\paragraph{}
The purpose of this article is to point out that every \emph{multi-monopole} solution of the BPS equations can be realised in a similar way, as a Berry phase in $\mathcal N = 4$ quantum mechanics. The key is to replace the twisted mass deformation with a new deformation, which, as far as the author is aware, has not previously appeared in the literature. This new deformation is constructed from matrix-valued functions of the moment map for a holomorphic isometry of $\mathbb{CP}^1$, and it preserves supersymmetry under the condition that these matrices obey a certain set of first-order equations.

\paragraph{}
In fact, these equations are nothing other than the Nahm equations. The computation of the Berry phase in the quantum-mechanical model with this  deformation turns out to be identical to the Nahm construction of BPS monopoles. The Nahm equations \cite{nahm, corrigan, hitchin, nakajima, donaldson} have already been studied in connection with many aspects of mathematical physics, including spectral curves \cite{hitchinspec}, D-branes \cite{diaconescu, hashimoto}, hyperk\"{a}hler geometry \cite{dancer}, geometric Langlands \cite{witten} and the vacuum geometry of theories in three or four dimensions \cite{gaiotto}. There is also a one-dimensional supersymmetric model \cite{fil} in which a version of the Nahm equations emerges as Dirac brackets in the quantum theory.

\paragraph{}
The present work offers a different perspective on the Nahm equations, showing that they are intimately related to K\"{a}hler geometry. While it may not immediately seem natural to view Nahm matrices as being defined over $\mathbb {CP}^1$ (as opposed to a real interval), this approach enables us to use the $\mathcal N = 4$ supersymmetry of quantum mechanics with K\"{a}hler target space in an elementary way to show that the gauge fields obtained from the Nahm construction must satisfy the BPS monopole equations.

\paragraph{}
Furthermore, our supersymmetric deformation generalises for any K\"{a}hler manifold with a holomorphic isometry. This leads to a procedure by which solutions of the BPS equations can be obtained from the zero modes of a certain deformation of the Dolbeault operators on any such K\"{a}hler manifold.

\section*{$\mathcal N = 4$ quantum mechanics and the Nahm equations}

\paragraph{}
We begin by describing a class of supersymmetric deformations of quantum mechanics on  K\"{a}hler target spaces with holomorphic isometry, which, as we will see, are closely related to the Nahm equations.

\paragraph{}
Let $\mathcal{M}$ be a complex manifold with a hermitian metric $g_{i\bar j}  = ( g_{ \bar i j})^\star $ such that $\omega = i g_{i \bar j} dz^i \wedge d\bar z^{\bar j} $ obeys  the K\"{a}hler condition,
\begin{eqnarray}
d \omega = 0 \nonumber
\end{eqnarray}
It is a standard fact that there exists a quantum-mechanical theory with target space $\mathcal{M}$ that is invariant under $\mathcal N = 4$ supersymmetry \cite{wittensusy}. The fields in the theory are a collection of chiral multiplets $(\phi^i, \psi^i)$, where $\phi^i$ are the complex coordinates on $\mathcal{M}$ and $\psi^i$ are two-component complex spinors in the holomorphic tangent bundle of $\mathcal{M}$.

\paragraph{}
Now suppose that $\mathcal{M}$ admits a holomorphic isometry, that is, there exists a global vector field $K = k^i \partial_i + \bar{k}^{\bar i} \partial_{\bar i}$, with $(k^i)^\star = \bar k^{\bar i}$, obeying the holomorphicity condition,
\begin{eqnarray}
\mathcal L_K J = 0 \nonumber
\end{eqnarray}
and the Killing equation,
\begin{eqnarray}
\mathcal L_K \omega = 0 \nonumber
\end{eqnarray}
where $J = idz^i \otimes \partial_i - i d\bar z^{\bar i} \otimes \partial_{\bar i}$ is the complex structure. Furthermore, suppose that there exists a global moment map $\mu$ for this isometry, obeying
\begin{eqnarray}
d\mu = \iota_K \omega \nonumber
\end{eqnarray}
It was shown in \cite{bagger} that one can gauge the isometry in a way that is invariant under $\mathcal N = 4$ supersymmetry. One introduces a vector multiplet $(u_t, x^a, \lambda, D)$, where $u_t$ is a $U(1)$ gauge field, $x^1, x^2, x^3$ are real scalars, $\lambda$ is a two-component complex spinor and $D$ is a real auxiliary scalar field. The Lagrangian is
\begin{eqnarray}
L = && \frac 1 {2e^2} \left( \dot x^a \dot x^a + i \bar \lambda \dot \lambda + D^2 \right)  + g_{\bar i j} D_t \bar\phi^{\bar i} D_t \phi^j - g_{\bar i j} \bar k^{\bar i} k^j  x^a x^a - \mu D \nonumber \\
&& +i g_{\bar i j} \bar {\psi}^{\bar i}D_t \psi^j - i \nabla_j k_{\bar i} x^a \bar{\psi}^{\bar i}  \sigma^a  \psi^j 
 - \frac 1 4 R_{i\bar k j \bar l} (\psi^{i T} \varepsilon \psi^j ) (\bar{\psi}^{\bar k} \varepsilon \bar{\psi}^{\bar l T})  - \bar k_i \psi^{i T} \varepsilon \lambda + k_{\bar i} \bar{\psi}^{\bar i} \varepsilon \bar \lambda^T \nonumber
\end{eqnarray}
Here, $\sigma^a$ are the Pauli matrices and $\varepsilon_{21} = - \varepsilon_{12} = +1$ is the alternating symbol. The covariant derivatives are defined as
\begin{eqnarray}
 D_t \phi^i = \dot \phi^i + u_t k^i, \ \ \ \ \ \ \ \ \ \ \ \  D_t \psi^i = \dot \psi^i + \Gamma^i_{jk} \dot \phi^j \psi^k + u_t \nabla_j k^i \psi^j \nonumber
\end{eqnarray}
and $\Gamma^j_{kl} = g^{j\bar m} \partial_k g_{\bar m l}$ and $R_{i \bar k j \bar l} = \partial_j \partial_{\bar l} g_{i \bar k} - g^{m \bar n} \partial_j g_{\bar n i} \partial_{\bar l} g_{m \bar k}$ are the standard expressions for the connection and curvature on a K\"{a}hler manifold. One can verify that this Lagrangian is invariant under the supersymmetry transformations
\begin{eqnarray}
\delta \phi^i & = & - \psi^{i T} \varepsilon \xi \nonumber \\
\delta \psi^i & = & \left( -i D_t \phi^i - i k^i   x^a \sigma^a   \right) \varepsilon \bar \xi^T  - \frac 1 2 \Gamma^i_{jk} ( \psi^{j T} \varepsilon \psi^k) \xi \nonumber \\
\delta u_t & = &  \frac i 2 \bar \lambda \xi - \frac i 2 \bar \xi \lambda \nonumber \\
\delta x^a & = &  \frac i 2 \bar \lambda \sigma^a \xi - \frac i 2 \bar \xi \sigma^a \lambda \nonumber \\
\delta \lambda & = & \dot x^a \sigma^a \xi - i D \xi \nonumber \\
\delta D & = & \frac 1 2 \bar \xi \dot \lambda + \frac 1 2 \dot{\bar \lambda } \xi
 \label{susytrans}
\end{eqnarray}

\paragraph{}
It is well-known that one can introduce an $\mathcal N = 4$ invariant mass term for the chiral multiplet fields. The construction is simple: we replace $x^a$ by $ x^a + m^a$, where $m^1, m^2, m^3$ are constants. The three mass parameters transform as a triplet under the $SU(2)_R$ symmetry. We refer to this deformation as a twisted mass, by analogy with terminology used for two-dimensional sigma models.  Of course, $x^a \mapsto x^a + m^a$ is a mere redefinition of variables, so as things stand, this deformation has no effect on the physics of the theory. However, if we send the gauge coupling $e$ to zero so that fluctuations of the fields in the vector multiplet cost infinite energy, and fix a particular zero-energy configuration for these fields, say $x^a = \lambda = D = 0$, then the parameters $m^a$ become genuine mass parameters.

\paragraph{}
One may ask whether it is possible to introduce a deformation similar to the twisted mass, but with the parameters $m^a$ replaced by functions that depend on the fields $\phi^i$ and $\bar \phi^{\bar i}$.  It turns out that such a construction exists if one is prepared to replace the parameters $m^a$ with a triplet of hermitian $k \times k$ \emph{matrices}, which we denote as $T^a$. To put matrices into a scalar-valued Lagrangian, we borrow a technique from \cite{gomis, me1, me2, ivanov, ivanov2}: we introduce a ``spin'' variable, a $k$-component bosonic vector $\varphi$ whose indices can be contracted with the indices on the matrices $T^a$. We give $\varphi$ a first-order kinetic term, and we gauge the symmetry $\varphi \mapsto e^{i \theta} \varphi$ by introducing a gauge field $\alpha_t$ that acts as a Lagrange multiplier imposing the constraint  $\varphi^\dagger \varphi = 1$.  Some experimentation reveals that a possible supersymmetric deformation of the theory  is
\begin{eqnarray}
L = &&  i \varphi^\dagger \dot \varphi + \alpha_t (\varphi^\dagger \varphi - 1) +  \frac 1 {2e^2} \left( \dot x^a \dot x^a + i \bar \lambda \dot \lambda + D^2 \right) \nonumber \\
&&  + g_{\bar i j} D_t \bar\phi^{\bar i} D_t \phi^j - g_{\bar i j} \bar k^{\bar i} k^j (  x^a + \varphi^\dagger T^a \varphi)(  x^a + \varphi^\dagger T^a \varphi) - \mu D \nonumber \\
&& +i g_{\bar i j} \bar {\psi}^{\bar i}D_t \psi^j - i \nabla_j k_{\bar i} ( x^a + \varphi^\dagger T^a \varphi )\bar{\psi}^{\bar i}  \sigma^a  \psi^j  - i k_{\bar i} \bar k_{j} \varepsilon^{abc} (\varphi^\dagger [T^b , T^c] \varphi) \bar{\psi}^{\bar i } \sigma^a \psi^j \nonumber \\
&&  - \frac 1 4 R_{i\bar k j \bar l} (\psi^{i T} \varepsilon \psi^j ) (\bar{\psi}^{\bar k} \varepsilon \bar{\psi}^{\bar l T})  - \bar k_i \psi^{i T} \varepsilon \lambda + k_{\bar i} \bar{\psi}^{\bar i} \varepsilon \bar \lambda^T \label{perturbedlagrangian}
\end{eqnarray}
where the hermitian matrices $T^a$ are functions of the moment map $\mu(\phi^i, \bar \phi^{\bar i})$, and obey the Nahm equation,
\begin{eqnarray}
\frac d {d\mu} T^a (\mu)  = \frac i 2 \varepsilon^{abc} [T^b (\mu) , T^c (\mu) ]  \label{nahm}
\end{eqnarray}
Indeed, whenever equation (\ref{nahm}) is satisfied, the variation of (\ref{perturbedlagrangian})  under the supersymmetry transformations (\ref{susytrans}) reduces to a total derivative upon imposing the equations of motion for $u_t$, $\varphi$ and $\psi^i$. By the standard Noether procedure, one obtains an expression for the conserved supercharges,
\begin{eqnarray}
Q = i g_{\bar i j} D_t \bar\phi^{\bar i} \psi^j + i \bar k_i (x^a +\varphi^\dagger T^a \varphi) \sigma^a \psi^i - \frac i 2 \mu \varepsilon \bar\lambda^T + \frac 1 {2e^2} \left( \dot x^a \sigma^a \varepsilon \bar\lambda^T + i D \varepsilon \bar \lambda^T \right) \nonumber
\end{eqnarray}

\paragraph{}
There are two key differences between the twisted mass deformation and the new deformation described. Firstly, as mentioned already,  the twisted mass only affects the physics of the theory in the limit $e \to 0$ in which the vector multiplet fields are frozen to a chosen supersymmetric configuration. The new deformation is different: it is also physically relevant for non-zero gauge coupling $e$.

\paragraph{}
Secondly, in the limit $e \to 0$, the twisted mass deformation remains supersymmetric if we do not include the gauge field $u_t$ in our theory; indeed, this is what is usually meant by a twisted mass deformation in the literature. (For instance, in the background $x^a = \lambda = D = 0$, the supersymmetry transformation for the fermion is modified to $\delta \psi^i = ( -i \dot \phi^i - i k^i   m^a \sigma^a   ) \varepsilon \bar \xi^T  - \frac 1 2 \Gamma^i_{jk} ( \psi^{j T} \varepsilon \psi^k) \xi$ and the supersymmetry algebra has a central charge.) Our new deformation, however, is only invariant under supersymmetry when the gauge field is present, even in the limit $e \to 0$, because the equation of motion for $u_t$ is necessary in order to ensure that the variation of the Lagrangian is a total derivative on-shell. We will return to this point in the next section, where we quantise the theory. 


\section*{Supersymmetric ground states and the associated Weyl equation}

\paragraph{}
In the Nahm construction of BPS monopoles, after a solution to the Nahm equations is obtained, the next step is to find zero modes of a certain Weyl operator built from this Nahm data. The purpose of this section is to explain how this step of the Nahm construction is related to a physical problem in our quantum-mechanical model: determining the supersymmetric ground states.

\paragraph{}
Throughout this section, we take the limit $e \to 0$. (In later sections, we will discuss what happens when we relax this condition.) As we have already discussed, this means that any fluctuations of the vector multiplet fields cost infinite energy, so the vector multiplet fields are frozen and take the form
\begin{eqnarray}
x^a = {\rm constant}, \ \  \ \ \lambda = 0, \ \ \ \ D = 0 \label{background}
\end{eqnarray}
We can think of $(x^1, x^2, x^3)$ as background parameters in the theory, while the chiral multiplet fields and the $k$-component field $\varphi$ remain dynamical. To quantise this theory, one must therefore quantise these dynamical fields in the supersymmetric background (\ref{background}).

\paragraph{}
Following Witten \cite{wittensusy}, canonical quantisation promotes the chiral multiplet fields to linear operators acting on differential forms on the K\"{a}hler manifold $\mathcal M$,
\begin{eqnarray}
\phi^i \mapsto z^i \times  \ \ \ \ \ \ \ \ \  & &  \ \ \ \ \ \ \ \ \bar{\phi}^{\bar i} \mapsto \bar {z}^{\bar i} \times  \nonumber
\\
g_{i \bar j} \dot{\bar{\phi}^{\bar j}} \mapsto - i \nabla_i \ \  \ \ \ & &  \ \  \ \  \ \  g_{\bar i j} \dot\phi^j \mapsto - i \nabla_{\bar i} \nonumber
\\
\psi^i \mapsto \left( \begin{array}{c}  dz^i \wedge \\  i g^{i \bar j} \iota_{\partial_{\bar j}} \end{array} \right) \ \ \ & & \ \ \  \bar \psi^{\bar i} \mapsto \left( \begin{array}{cc} g^{\bar i j} \iota_{\partial_j} & - i d\bar z^{\bar i} \wedge \end{array} \right) \label{dictionary}
\end{eqnarray}
Meanwhile, the canonical commutation relations for the $k$-component bosonic variable $\varphi_p$ are
\begin{eqnarray}
[\varphi_p, \varphi^\dagger_q] = \delta_{pq} \nonumber
\end{eqnarray}
We define a state $| 0 \rangle $ such that $ \varphi_p | 0 \rangle = 0  $ for $p = 1, ... , k$. Observe that the equation of motion obtained by varying the Lagrange multiplier field $\alpha_t$ imposes the constraint $ \sum_{p = 1}^k \varphi_p^\dagger \varphi_p = 1 $. Therefore, the allowed states in the theory are of the form
\begin{eqnarray}
\sum_{p = 1}^k \eta_p \varphi^\dagger_p |0\rangle \nonumber
\end{eqnarray}
where $\eta_1, ... , \eta_k$ are differential forms on $\mathcal M$. Collecting these differential forms into a $k$-component column vector $\eta = (\eta_1, ... , \eta_k )^T$, we see that the operator $\varphi^\dagger T^a \varphi$ acts  on $\eta$ as multiplication by the matrix $T^a(z^i, \bar z^{\bar i})$.

\paragraph{}
Since the  theory is a $U(1)$ gauge theory, we must fix a gauge when canonically quantising. We pick the $u_t = 0$ gauge. In this gauge the equation of motion for $u_t$ reads as $i\bar k_i \dot \phi^i + i k_{\bar i} \dot{\bar \phi^{\bar i}} - \nabla_j k_{\bar i} \bar{\psi}^{\bar i} \psi^j = 0$, and this equation must be imposed as a constraint on the Hilbert space. By the dictionary (\ref{dictionary}), the physical states in the Hilbert space are those differential forms that are  invariant under the action of the isometry, that is, those that are annihilated by the Lie derivative with respect to the Killing vector field $K = k^i \partial_i + \bar k^{\bar i} \partial_{\bar i}$,
\begin{eqnarray}
\mathcal L_K \eta  = 0  \label{physicalstates}
\end{eqnarray} 

\paragraph{}
To complete the quantisation procedure, we must specify the inner product on the states. This is  given by the usual expression,
\begin{eqnarray}
\langle \eta ' | \eta \rangle = \sum_{p = 1}^k \int \eta_p \wedge \star \eta_p' \nonumber
\end{eqnarray} 

\paragraph{}
We are now almost ready to write down the supercharges as differential operators. Before doing so,  it helps to organise the various Dolbeault operators and contraction operators as doublets of the Lefshetz $SU(2)$ action on the K\"{a}hler manifold $\mathcal M$, which descends from the $SU(2)_R$ action in the quantum mechanics. We define
\begin{eqnarray}
d_\alpha = \left( \begin{array}{c} \partial \\ - i \bar \partial^\dagger \end{array} \right) \ \ \ \ \bar d_\alpha = \left( \begin{array}{cc} \partial^\dagger & i \bar \partial \end{array} \right) \nonumber
\end{eqnarray}
\begin{eqnarray}
\iota_\alpha = \left( \begin{array}{c} i  \iota_{(k^i \partial_i)}^\dagger \\ - \iota_{(\bar k^{\bar i} \partial_{\bar i}) }\end{array} \right) \ \ \ \ \bar \iota_\alpha = \left( \begin{array}{cc} -i   \iota_{(k^i \partial_i)}  & \  -\iota_{(\bar k^{\bar i} \partial_{\bar i}  )}^\dagger \end{array} \right) \nonumber
\end{eqnarray}
The K\"{a}hler condition, Killing's equation, holomorphicity of the Killing field and the Nahm equation (\ref{nahm}) can be summarised as a list of (anti)commutation relations,
\begin{eqnarray}
\{ d_\alpha, d_\beta \} = \{ \bar d_\alpha, \bar d_\beta \} = 0, \ \  \ \  \{d_\alpha,  \bar d_\beta \} = \Delta \delta_{\alpha \beta} = & \frac 1 2 (d d^\dagger + d^\dagger d) \delta_{\alpha \beta} \nonumber
\end{eqnarray}
\begin{eqnarray}
\{ \iota_\alpha, \iota_\beta \} = \{ \bar \iota_\alpha , \bar \iota_\beta \} = 0 , \ \ \ \ \{ \iota_\alpha , \bar \iota_\beta \} = & \frac 1 2 \sum_\gamma \{ \iota_\gamma, \bar \iota_\gamma \} \delta_{\alpha \beta} \nonumber
\end{eqnarray}
\begin{eqnarray}
\{ d_\alpha , \iota_\beta \} = \{ \bar d_\alpha, \bar \iota_\beta \} = 0 , \ \ \  \ \ \  \{ d_\alpha, \bar \iota_\beta \} - \{ \iota_\alpha , \bar d _\beta \} = & \frac 1 2  \sum_{\gamma} \left( \{ d_\gamma , \bar \iota _\gamma \} - \{ \iota_\gamma , \bar d_\gamma \} \right) \delta_{\alpha \beta} \nonumber
\end{eqnarray}
\begin{eqnarray}
\mathcal L_K = & \frac i 2 \sum_{\gamma} \left( \{ d_\gamma , \bar \iota_\gamma \} + \{ \iota_\gamma , \bar d_\gamma \} \right) \nonumber
\end{eqnarray}
\begin{eqnarray}
[d_\alpha , T^a ] = -  & \frac { i }{2} \varepsilon^{abc} [T^b, T^c] \iota_\alpha, \ \  \ \ [\bar d_\alpha , T^a ] = +  \frac i 2 \varepsilon^{abc} [T^b, T^c ] \bar \iota_\alpha \nonumber 
\end{eqnarray}
\begin{eqnarray}
[\iota_\alpha,  T^a ] = [\bar \iota_\alpha, T^a ] = 0  \label{relations}
\end{eqnarray}

\paragraph{}
Having established this notation, we find the following expressions for the supercharges,
\begin{eqnarray}
 Q_\alpha = d_\alpha +(x^a +  T^a)    \sigma^a_{\alpha \beta } \iota_\beta, \ \ \ \ \bar Q_\alpha = \bar d_\alpha + (x^a + T^a )   \bar \iota_\beta \sigma_{\beta \alpha} \label{supercharges}
\end{eqnarray}
The Hamiltonian is
\begin{eqnarray}
& H =  \Delta   +  \frac 1 2  (x^a + T^a)(x^a + T^a )  \{ \iota_\gamma , \bar \iota_\gamma \} + (x^a + T^a) \sigma_{\gamma \delta}^a \{ \iota_\delta, \bar d_\gamma\} - i \varepsilon^{abc} [T^b, T^c ] \sigma^a_{\gamma \delta }  \iota_\delta \bar \iota_\gamma \nonumber
\end{eqnarray}
One can check that the list of relations (\ref{relations}) ensures that the supercharges and Hamiltonian preserve the space of physical states (\ref{physicalstates}), and, when acting on this space, obey the supersymmetry algebra
\begin{eqnarray}
\{ Q_\alpha, Q_\beta \} = \{ \bar Q_\alpha , \bar Q_\beta \} = [H, Q_\alpha] = [H, \bar Q_\alpha ] =  0, \ \ \ \ \ \{ Q_\alpha , \bar Q_\beta \} = H \delta_{\alpha \beta}  \nonumber
\end{eqnarray}
Note that the gauge fixing constraint $\mathcal L_K \eta = 0$ is essential to obtain the commutator $ [H, Q_\alpha] = [H, \bar Q_\alpha ] =  0$. (By contrast, when one uses constant, scalar-valued twisted masses $(m^1, m^2, m^3)$,  the commutation relations $ [H, Q_\alpha] = [H, \bar Q_\alpha ] =  0$ hold regardless of the constraint, a sign that the theory remains supersymmetric even without gauging the isometry; in this case the $Q_\alpha$ and $\bar Q_\beta$ anticommutator becomes $\{ Q_\alpha , \bar Q_\beta \} = H \delta_{\alpha \beta} - i \mathcal L_K m^a \sigma^a_{\alpha \beta}$, and $\mathcal L_K$ is interpreted as a central charge. This is the setup considered in \cite{sonner1}.)

\paragraph{}
One may notice a similarity between the supercharges $Q_\alpha$ and $\bar Q_\alpha$ in (\ref{supercharges}) and the Weyl operators used in the Nahm construction of monopoles. Zero modes of these supercharges have a physical interpretation: they represent the ground states of energy $H=0$. 

\paragraph{}
In one special example, the supercharges reduce precisely to the Weyl operators familiar from the Nahm construction. This is when $\mathcal {M}$ is the Riemann sphere $\mathbb{CP}^1$. We use  a stereographic coordinate $z$ on  $ \mathbb{CP}^1 \backslash \{ \infty \} $, chosen so that the isometry is the rotation $z \mapsto e^{-i\theta} z$. Then the metric is of the form $g_{z \bar z} = g\left( |z |^2 \right)$ for some positive function $g$, and the Killing vector field associated to the rotation is $-iz \partial_z + i \bar z \partial_{\bar z}$. The moment map for the isometry is $\mu = \mu \left(| z |^2\right)$, where $\mu' =  g$; it is a monotonically increasing function of $| z|$, taking values in the interval $[ \mu(0), \mu(\infty)]$. 

\paragraph{}
To see that the supercharges on $\mathcal M = \mathbb {CP}^1$ really do reduce to the familiar Weyl operators, we first observe that the operators $ \sigma^a_{\gamma \delta } \{ \iota_\delta, \bar d_\gamma\} $ and $ \sigma^a_{\gamma \delta }\iota_\delta  \bar \iota_\gamma $ that appear in the Hamiltonian vanish on (1,0)- and (0,1)-forms, and what remains of the Hamiltonian is positive definite because $ h^{1,0} (\mathbb {CP}^1) = h^{0,1} (\mathbb {CP}^1) = 0$ while  $\frac 1 2  (x^a + T^a) (x^a + T^a ) \{ \iota_\gamma , \bar \iota_\gamma \}  =  (x^a + T^a) (x^a + T^a ) |z|^2 g\left(| z |^2 \right) $ is the square of a hermitian matrix. So for the purposes of finding supersymmetric ground states, we may restrict our attention to linear combinations of (0,0)- and (1,1)-forms. Let us write such a state as
\begin{eqnarray}
\eta = \frac 1 {\sqrt{2\pi} } \left( f_1 + f_2 \omega \right) \nonumber
\end{eqnarray}
where $f_1$ and $f_2$ are $k$-component vector-valued functions on $\mathbb{CP}^1$ and $\omega = i g_{z\bar z} dz \wedge d \bar z$ is the K\"{a}hler form. The gauge fixing constraint (\ref{physicalstates}) implies that $f_1, f_2$ are functions of the moment map $\mu$. The condition $Q_\alpha \eta = \bar Q_\alpha \eta = 0 $ defining supersymmetric ground states is equivalent to
\begin{eqnarray}
\left( \frac d {d\mu} - \left( x^a 1_{k\times k}+ T^a(\mu) \right) \otimes \sigma^a \right) \left(  \begin{array}{c} f_1 (\mu)  \\ f_2 (\mu) \end{array} \right) = 0 \label{weyl}
\end{eqnarray}
which is precisely the Weyl equation associated with the Nahm data $T^a(\mu)$.

\section*{Berry's connection and the inverse Nahm transform}

\paragraph{}
In the previous section, we described the supersymmetric ground states in our theory in the limit $e \to 0$ in which the vector multiplets are frozen. We now consider what happens when the gauge coupling $e$ is small but non-zero, so that $x^a $ may vary slowly with time. We shall see that the quantum dynamics is governed by a Berry connection, and the construction of this Berry connection is related to the last step of the Nahm construction in which solutions of the BPS monopole equations are finally obtained.

\paragraph{}
We will continue to think of $x^a$ as a background parameter, but now as one that varies slowly with time. The supersymmetric ground states are still represented by differential forms annihilated by the supercharge operators given in (\ref{supercharges}), but since $x^a$ appears as a parameter in these supercharge operators, the states will change adiabatically as $x^a$ varies. If $\mathcal M$ is compact, the spectrum of the Hamiltonian is discrete, and as $x^a$ varies, the number of supersymmetric ground states can only ever change by an even number. This is because, for every excited state, there is at least one supercharge, say $Q_\alpha$, that fails to annihilate that state. So the excited states in the theory come in pairs, $\{|\eta\rangle , Q_\alpha | \eta \rangle \}$. Thus the Witten index $I = {\rm Tr} (-1)^F$ is preserved as $x^a$ varies, and the number of supersymmetric ground states always remains equal to the Witten index modulo 2.

\paragraph{}
Let us make the assumption that, in the particular quantum-mechanical model we are considering, the number of supersymmetric ground states is exactly equal to $| I |$ for all $x^a$. (We will soon comment on what happens if this condition is violated.) We pick a basis of ground states,  $\{ |m (\vec x) \rangle\} $, with $m= 1, ... , |I|$, obeying the orthonormality condition
\begin{eqnarray}
\langle m (\vec x) | n (\vec x) \rangle= \delta_{mn} \nonumber
\end{eqnarray}
Writing a general state as a linear combination,
\begin{eqnarray}
\sum_m \chi_m (t) | m (\vec x (t)) \rangle \nonumber
\end{eqnarray}
where $\chi_m(t)$ are complex coefficients with $\sum_m \chi^\star_m \chi_m = 1$, it is easy to see that the Schr\"{o}dinger equation,
\begin{eqnarray}
i \partial_t \left( \sum_m \chi_m (t) | m (\vec x(t)) \rangle \right) = \hat H \left(  \sum_m \chi_m (t) | m (\vec x(t)) \rangle \right) = 0 \nonumber
\end{eqnarray}
implies that the coefficients $\chi_m(t)$ satisfy the first-order equation of motion,
\begin{eqnarray}
\dot \chi_m = i \dot x^a  \sum_{n} A_a (\vec x)_{mn} \chi_n \label{timeevolution}
\end{eqnarray}
where
\begin{eqnarray}
A_a(\vec x)_{mn} = i \langle m(\vec x) | \frac \partial {\partial x^a } |  n(\vec x) \rangle \label{gaugefield}
\end{eqnarray}
The hermitian matrix $A_a(\vec x)_{mn}$ is the \emph{Berry connection} for our model. Geometrically, we can think of $(x^1, x^2, x^3)$ as coordinates on a parameter space $\mathbb R^3$, and we can view $\chi_m$ as coordinates on the fibres of a complex vector bundle of rank $|I|$ over this $\mathbb R^3$. This bundle is known as the vacuum bundle. Equation (\ref{timeevolution}) is the condition of parallel transport with respect to the connection $A_a$. The Berry connection $A_a(\vec x)$ is a smooth connection; however, if we relax the condition that the number of supersymmetric ground states equals $| I|$ for all $(x^1,x^2,x^3)$, that is, if extra pairs of ground states appear at certain points in the parameter space $\mathbb R^3$, then the vacuum bundle is ill-defined at those points in the parameter space and the Berry connection develops singularities at those points.

\paragraph{}
Let us return to our special example where $\mathcal M $ is a rotationally invariant $\mathbb{CP}^1$. Here, the definition of the Berry connection (\ref{gaugefield}) is identical to the standard construction of the Yang-Mills fields in a BPS monopole from Nahm data,
\begin{eqnarray}
A_a (\vec x)_{mn} =  i  \int_{\mu(0)}^{\mu(\infty)} d\mu \left( f_{m,1}^\dagger ( \vec x, \mu) \frac \partial {\partial x^a } f_{n,1} (\vec x, \mu)+ f_{m,2}^\dagger (\vec x , \mu) \frac \partial {\partial x^a }  f_{n,2} (\vec x, \mu) \right) \nonumber
\end{eqnarray}
The rank of the vacuum bundle is equal to the number of supersymmetric ground states, and this depends on the precise form of the Nahm data. In \cite{corrigan}, it was shown by an index theorem \cite{callias} that if the matrices $T^a$ have simple poles at $\mu = \mu(0)$ and $\mu(\infty)$, with residues defining the $k$-dimensional irreducible representation of $su(2)$, then there are  two solutions to the Weyl equation (\ref{weyl}) with finite $L^2$-norm, that is, there are  two  supersymmetric ground states. Thus, the Berry connection (\ref{gaugefield}) is a $U(2)$ connection. If we rescale $\mu$ by an additive constant so that $\mu(0) = - \mu(\infty) $, and impose the condition $T^t(-\mu) = T(\mu)$, the Berry connection is also an $SU(2)$ connection \cite{corrigan}.

\paragraph{}
So far, we have only identified the Yang-Mills fields in the BPS monopole solution. It still remains to identify the Higgs field. This is done by defining an endomorphism $\Phi_{mn} (\vec x)$ on the vacuum bundle to be the matrix elements of the moment map,
\begin{eqnarray}
\Phi(\vec x)_{mn}  = - \langle m(\vec x) |  \mu (\phi^i, \bar \phi^{\bar i}) | n(\vec x) \rangle \label{higgsfield}
\end{eqnarray}
We postpone the physical interpretation of $\Phi_{mn}$ for the next section.

\paragraph{}
In the example where  $\mathcal M = \mathbb {CP}^1$, this coincides with the usual construction of the Higgs field from Nahm data,
\begin{eqnarray}
\Phi (\vec x)_{mn} =  -  \int_{\mu(0)}^{\mu(\infty)} d\mu \left( \mu f_{m,1}^\dagger ( \vec x, \mu)  f_{n,1} (\vec x, \mu)+ \mu f_{m,2}^\dagger (\vec x , \mu)  f_{n,2} (\vec x, \mu) \right) \nonumber
\end{eqnarray}
It is a standard result that the $SU(2)$ connection $A_a$ and endomorphism $\Phi$ obtained in this way obey the BPS monopole equation $D_a \Phi = \frac 1 2 \varepsilon^{abc} F_{bc}$. In fact, there is a one-one correspondence between Nahm data and BPS monopole solutions \cite{hitchin, nakajima}. The situation where the Nahm matrices are $1 \times1$ matrices (i.e. constant twisted masses) and the resulting BPS monopole solution has unit magnetic charge was studied by Sonner and Tong in \cite{sonner1, sonner2}. What we have done is generalise the procedure of Sonner and Tong for Nahm matrices of arbitrary size, showing that the construction produces all multi-monopole solutions of the BPS equations.

\paragraph{}
Note that the role of the pole condition described above is only to ensure that there are two supersymmetric ground states for all values of $x^a$, so the vacuum bundle is of rank two. The BPS equation $D_a \Phi = \frac 1 2 \varepsilon^{abc} F_{bc}$, however, holds regardless of the residues of the poles of $T^a$ at $\mu = \mu(0)$ and $\mu(\infty)$. For example, if we set $T^a$ to be $2\times 2$ zero matrices on $\mathbb {CP}^1$, so the residues of the poles of $T^a$ at $\mu = \mu(0)$ and $ \mu(\infty)$ are direct sums of two copies of the 1-dimensional trivial representation of $su(2)$, then the system has four supersymmetric ground states, and $A_a$ and $\Phi$ are an $SU(2) \times SU(2)$ connection and endomorphism on a rank four complex vector bundle over $\mathbb R^3$, equal to the direct sum of two $SU(2)$ monopoles of unit magnetic charge.

\paragraph{}
The BPS equation also holds regardless of the choice of the K\"{a}hler manifold $\mathcal M$ (although for general $\mathcal M$ there is no guarantee that $A_a$ and $\Phi$ will be non-singular); demonstrating this will be the goal of the next section.

\section*{The effective action and the BPS monopole equations}

\paragraph{}
In this final section, we integrate out the fast degrees of freedom in our model, that is, we integrate out the fields in the chiral multiplets, leaving an effective action for the fields in the vector multiplet. We will see that both the Berry connection $A_a(\vec x)_{mn}$ and the endomorphism $\Phi(\vec x)_{mn}$ defined in the previous section will appear in this effective action, and the requirement that this effective action is invariant under $\mathcal N = 4$ supersymmetry forces $A_a$ and $\Phi$ to obey the BPS monopole equations. Our presentation here roughly follows \cite{sonner2}, though the idea of using supersymmetry to constrain connection terms in an effective action can be traced back to \cite{cecotti, smilga, denef}. (In fact, the result in \cite{smilga, denef} is a special case of the result presented here, with $\mathcal M = \mathbb C$ and $T^a$ equal to the $1\times 1$ zero matrices; in this setup, the vacuum bundle is a rank one vector bundle, singular at the origin in $\mathbb R^3$, and the connection $A_a$ is the Dirac monopole, while the endomorphism $\Phi$ obeys $\partial_a \Phi = \frac 1 2 \varepsilon^{abc} F_{bc}$.) 

\paragraph{}
A subtlety in the computation of this effective action is that the system has multiple, degenerate, ground states. Therefore, the effective action depends not only on the vector multiplet fields, but also on the variable $\chi_m(t)$ that indicates which ground state the system is in at any moment in time. (Alternatively, one can present the effective action as a matrix-valued weight in the path integral; this equivalent approach was taken in \cite{sonner2}.)

\paragraph{}
One can check by dimensional analysis \cite{denef} that the effective action contains two types of bosonic terms at lowest order in the derivative expansion: a term proportional to $\dot x^a$, and another proportional to $D$. The first of these terms can be deduced from the discussion in the previous section. We already know that, as $x^a$ varies over time, $\chi_m$ obeys the equation of motion (\ref{timeevolution}). The action that gives rise to such an equation of motion is
\begin{eqnarray}
L_{\dot{\vec x}} = i \chi^\dagger \dot \chi  +\beta_t (\chi^\dagger \chi - 1 ) + \dot x^a  \left( \chi^\dagger A_a(\vec x) \chi \right) \nonumber
\end{eqnarray}
and this provides the term in the effective action proportional to $\dot x^a$. (Note that we have introduced an additional variable $\beta_t$, which acts as a Lagrange multiplier imposing the normalisation $\chi^\dagger \chi = 1$; we used a similar trick with $\varphi$ and $\alpha_t$ when we constructed the original Lagrangian, equation (\ref{perturbedlagrangian}).)

\paragraph{}
Next, rather than allowing $x^a$ to vary, we consider instead what happens if we allow $D$ to acquire a non-zero, but still time-independent, vacuum expectation value. By examining the original Lagrangian, we learn that turning on a vacuum expectation value for $D$ shifts the energy of the ground states by
\begin{eqnarray}
\Delta H = D \langle  \mu \rangle  \nonumber
\end{eqnarray}
This energy shift can be written in terms of the endomorphism $\Phi_{mn}(\vec x)$ defined in (\ref{higgsfield}) as
\begin{eqnarray}
\Delta H = - D \left(\chi^\dagger \Phi(\vec x) \chi  \right) \nonumber
\end{eqnarray}
It follows that the second of the terms in the effective action, the term depending on $D$, is
\begin{eqnarray}
L_{D} = +  D \left( \chi^\dagger \Phi(\vec x) \chi \right) \nonumber
\end{eqnarray}

\paragraph{}
All that remains is to find the fermionic terms in the effective action. These are constrained by the requirement that the effective action must be invariant under the supersymmetry transformations for $x^a$, $\lambda$ and $D$ in (\ref{susytrans}). As shown in \cite{sonner2}, a supersymmetric completion of $L_{\dot{\vec x}} + L_D$ only exists if $A_a$ and $\Phi$ obey
\begin{eqnarray}
D_a \Phi = \frac 1 2 \varepsilon^{abc} F_{bc} \label{BPS}
\end{eqnarray} 
where
\begin{eqnarray}
D_a \Phi = \frac \partial {\partial x^a} \Phi - i [A_a, \Phi ], \ \ \ \ \ \ F_{bc} =   \frac \partial {\partial x^b}  A_c - \frac \partial {\partial x^c} A_b - i [A_b , A_c]   \nonumber
\end{eqnarray}
and, provided equation (\ref{BPS}) is obeyed, the unique possibility for the effective action is
\begin{eqnarray}
L   =  &&  \frac 1 {2e^2}  \left( \dot x^a \dot x^a + i \bar \lambda \dot \lambda +  D^2 \right)
  + i \chi^\dagger \dot \chi  +\beta_t (\chi^\dagger \chi - 1 ) \nonumber \\ && + \dot x^a \left( \chi^\dagger A_a \chi \right)+ \frac 1 2 (\bar \lambda \sigma^a \lambda ) \left( \chi^\dagger D_a \Phi \chi \right) + D \left(\chi^\dagger \Phi \chi \right) \nonumber
\end{eqnarray} 
Of course, equation (\ref{BPS}) is the BPS monopole equation.

\paragraph{}
It is pleasing that both the Nahm equation and the BPS equation drop out of our quantum-mechanical model as conditions for $\mathcal N = 4$ supersymmetry: the Nahm equation ensures that our matrix-valued version of the twisted mass deformation is supersymmetric, while the BPS equation does the same for the effective action for the vector multiplet. To impose $\mathcal N = 4$ supersymmetry, it is necessary to think of the Nahm matrices as being defined over a K\"{a}hler manifold $\mathcal M$ with a holomorphic isometry, rather than on a real interval; the Nahm matrices then form a triplet under the Lefschetz $SU(2)_R$ action. When $\mathcal M$ is $\mathbb {CP}^1$, the steps towards the computation of the effective action are in correspondence with the steps in the standard Nahm construction of BPS monopoles. But the arguments presented here imply that the procedure is more general: starting from any arbitrary K\"{a}hler manifold with a holomorphic isometry, and from any solution of the Nahm equations, there is a canonical way to construct a (possibly singular) connection and endomorphism on a complex vector bundle over $\mathbb R^3$ obeying the BPS monopole equations. 

\section*{Acknowledgements}
The author would like to thank David Tong for a careful reading of this article. The author is supported by Gonville and Caius College and the ERC Grant agreement STG 279943.


\begin{thebibliography}{1}

\bibitem{berry} M. V. Berry, \emph{``Quantal Phase Factors Accompanying Adiabatic Changes,''} Proc. R. Soc. A \textbf{392} (1984) 45.

\bibitem{sonner1} J. Sonner and D. Tong, \emph{``Non-Abelian Berry Phases and BPS Monopoles,''} Phys. Rev. Lett. \textbf{102}  (2009) 191801, arXiv:0809.3783.

\bibitem{thooft} G. 't Hooft, \emph{``Magnetic Monopoles in Unified Gauge Theories,''} Nucl. Phys. B \textbf{79} (1974) 276.

\bibitem{polyakov} A. M. Polyakov, \emph{``Particle Spectrum in Quantum Field Theory,''} JETP \textbf{20} (1974) 194.

\bibitem{sonner2} J. Sonner and D. Tong, \emph{``Berry Phase and Supersymmetry,''} JHEP \textbf{0901} (2009) 063, arXiv:0810.1280.

\bibitem{bogomolny} E. B. Bogomolny, \emph{``The Stability of Classical Solutions,''} Sov. J. Nucl. Phys. \textbf{24}  (1976) 49.

\bibitem{ps} M. K. Prasad and C. M. Sommerfield, \emph{``An Exact Classical Solution for the  't Hooft Monopole and the Julia-Zee Dyon,''} Phys. Rev. Lett. \textbf{35} (1975) 760.

\bibitem{cecotti} S. Cecotti and C. Vafa, \emph{``Topological-Antitopological Fusion,''} Nucl. Phys. B \textbf{367} (1991) 359.

\bibitem{smilga} A. Smilga, \emph{``Vacuum Structure in the Chiral Supersymmetric Quantum Electrodynamics,''} JETP \textbf{64} (1986) 8.

\bibitem{denef} F. Denef, \emph{``Quantum Quivers and Hall/Hole Halos,''} JHEP \textbf{0210} (2002) 023,
arXiv:hep-th/0206072.

\bibitem{nahm} W. Nahm, \emph{``A Simple Formalism For The BPS Monopole,''} Phys. Lett. B \textbf{90} (1980) 413.

\bibitem{corrigan} E. Corrigan and P. Goddard, \emph{``Construction of Instanton and Monopole Solutions and Reciprocity,''} Ann. Phys. \textbf{154} (1984) 253.

\bibitem{hitchin} N. Hitchin, \emph{``On the Construction of Monopoles,''} Commun. Math. Phys. \textbf{89}  (1983) 145.

\bibitem{nakajima} H. Nakajima, \emph{``Monopoles and Nahm's Equations,''} in  Einstein Metrics and Yang-Mills Connections, ed. T.  Mabuchi and S. Mukai (1990) 193.

\bibitem{donaldson} S. K. Donaldson, \emph{``Nahm's Equations and the Classification of Monopoles,''} Commun. Math. Phys. \textbf{96} (1984) 387.

\bibitem{hitchinspec} N. Hitchin, \emph{``Monopoles and Geodesics,''} Commun. Math. Phys. \textbf{83} (1982) 579.

\bibitem{diaconescu} D.-E. Diaconescu, \emph{``D-branes, Monopoles and Nahm Equations,''} Nucl. Phys. B \textbf{503} (1997) 220,  arXiv:hep-th/9608163.

\bibitem{hashimoto} K. Hashimoto and S. Terashima, \emph{``Stringy Derivation of Nahm Construction of Monopoles,''} JHEP \textbf{0509} (2005) 055, arXiv:hep-th/0507078.

\bibitem{dancer} A. Dancer, \emph{``Nahm's Equations and Hyperk\"{a}hler Geometry,''} Commun. Math. Phys. \textbf{158} (1993) 545.

\bibitem{witten} E. Witten, \emph{``Geometric Langlands and the Equations of Nahm and Bogomolny,''} arXiv:0905.4795.

\bibitem{gaiotto} S. Cecotti, D. Gaiotto and C. Vafa, \emph{``tt$^\star$ Geometry in 3 and 4 Dimensions,''} JHEP \textbf{1405} (2014) 055, arXiv:1312.1008. 

\bibitem{fil} S. Fedoruk, E. Ivanov and O. Lechtenfeld, \emph{``Nahm Equations in Supersymmetric Mechanics,''} JHEP \textbf{1206} (2012) 147, arXiv:1204.4474.

\bibitem{wittensusy} E. Witten, \emph{``Constraints on Supersymmetry Breaking,''} Nucl. Phys. B \textbf{202} (1982) 253.

\bibitem{bagger} J. Bagger and E. Witten, \emph{``The Gauge Invariant Supersymmetric Nonlinear Sigma Model,''} Phys. Lett. B \textbf{118} (1982) 103.

\bibitem{gomis} J. Gomis and F. Passerini, \emph{``Wilson Loops as D3-Branes,''} JHEP \textbf{0701} (2007) 097, arXiv:hep-th/0612022. 

\bibitem{me1} D. Tong and K. Wong, \emph{``Monopoles and Wilson Lines,''} JHEP \textbf{1406} (2014) 048, arXiv:1401.6167.

\bibitem{me2} D. Tong and K. Wong, \emph{``ADHM Revisited: Instantons and Wilson Lines,''} Phys. Rev. D \textbf{91} (2015) 026007, arXiv:1410.8523.

\bibitem{ivanov} E. Ivanov, M. Konyushikhin and A. Smilga, \emph{``SQM with Non-Abelian Self-Dual Fields: Harmonic Superspace Description,''} JHEP \textbf{1005} (2010) 033, arXiv:0912.3289. 

\bibitem{ivanov2} E. Ivanov and M. Konyushikhin, \emph{``N = 4, 3D Supersymmetric Quantum Mechanics in Non-Abelian Monopole Background,''} Phys. Rev. D \textbf{82} (2010) 085014,  arXiv:1004.4597.

\bibitem{callias} C. Callias, \emph{``Index Theorems on Open Spaces,''}  Commun. Math. Phys. \textbf{62} (1978) 213.


\end{thebibliography}
\end{document}